\input{aipcheck}

\documentclass[,final] {aipproc}

\layoutstyle{6x9} \usepackage{amssymb}

\newcommand{\beq}{\begin{equation}} \newcommand{\eeq}{\end{equation}}

\newcommand{\Neq}{N_{\rm eq}}

\newcommand{\deltat}{\deltat}

\newcommand{\mchi}{m_{\chi}}

\newcommand{\Cc}{C_\mathrm{c}} 
\newcommand{\Ca}{C_\mathrm{a}}

 \def\yr {\ifmmode \,\, {\rm yr} \else yr \fi}

\begin{document}

\title{Energetic neutrinos from the Sun and Earth and dark matter
substructure}


\author{Savvas M. Koushiappas}{ address={Department of Physics, Brown
University, 182 Hope St., Providence, RI 02912, U.S.A.}  }

\author{Marc Kamionkowski}{ address={California Institute of
Technology, Mail Code 350-17, Pasadena, CA 91125, U.S. A.}  }

%---------------- ABSTRACT
\begin{abstract} Dark matter halos contain a wealth of substructure in
the form of subhalos and tidal streams. Enhancements in the dark
matter density of these regions leads to enhanced rates in direct
detection experiments, as well as enhanced dark matter capture rates
in the Sun and the Earth. Direct detection experiments probe the
present-day dark matter density, while energetic neutrinos probe
the past history of the dark matter density along the solar system's orbit 
about the Galactic center \cite{Koushiappas:2009ee}. We discuss how an elevated 
energetic neutrino flux can be used to probe the level of substructure
present at the Galactic radius of the solar system.
\end{abstract}

\maketitle

%---------------- INTRODUCTION

\section{Introduction}

One of the experimental probes of the nature of dark matter is the
possible detection of high-energy neutrinos from the Sun and the Earth 
\cite{neutrinos,Gould:1987ir}. If the dark matter particle is a Weakly
Interacting Massive Particle (WIMP)
\cite{Jungman:1995df,Hooper:2007qk}, then interactions between WIMPs
in the Milky Way halo and nuclei in the Sun/Earth can reduce the speed of
the WIMPs so that they get captured by the gravitational potential
well of the Sun/Earth. As the accumulation of WIMPs increases in the solar
interior, so does their annihilation rate. Final state neutrinos from
WIMP annihilation escape the Sun, and travel freely to the Earth (similarly, 
neutrinos from the center of the Earth travel to the surface) where
they may get detected in a high energy neutrino experiment such as
IceCube \cite{DeYoung:2009er}. The expected rate of high energy
neutrinos depends on the rate at which WIMPs are gravitationally
captured (i.e., the dark matter density), and the rate at which they
are depleted by annihilation.

Direct detection of dark matter (e.g. in a liquid noble gas
\cite{lux}) probes the present day dark matter density at the solar
galactic radius. Therefore, the rate in a direct detection experiment
is proportional to the rate of high-energy neutrinos from the Sun/Earth 
\cite{Kamionkowski:1994dp}. However, the proportionality is time
independent only if the dark matter density is time independent. The
presence of Galactic substructure
\cite{substructure,Kamionkowski:2008vw,earthmass} suggests that the
constant of proportionality can change with time, depending on the {\it past
history} of the dark matter density along the Sun's orbit about the
Galactic center. As such, it may be possible
to probe Galactic substructure using future measurements of the flux
of energetic neutrinos from the Sun/Earth and direct detection
experiments  \cite{Koushiappas:2009ee}. In this proceedings we summarize these effects, and
discuss how future measurements can be used to probe Galactic
substructure.

% -------------- THE STANDARD CASE
\section{Energetic solar neutrinos from the Sun and Earth}

\subsection{The classical case}

In the standard scenario, the flux of upward muons from WIMP
annihilation in the Sun is
\begin{eqnarray}
\label{eqn:neutrinos} \Gamma_{\nu,0} &=& 7.3\times10^5\, {\rm
km}^{-2}\, {\rm yr}^{-1}\, \left( \frac{N}{\Neq}\right)^2
\left(\frac{\rho_{\chi}}{0.3 {\rm GeV cm}^{-3}} \right)
\left(\frac{\mchi}{100 {\rm GeV} } \right)^2 \nonumber \\ 
&\times&
\left(\frac{\sigma}{10^{-40} {\rm cm}^{2}} \right)
\left[\frac{\xi(\mchi)}{0.1}\right] f(\mchi) \,
\end{eqnarray}
while the corresponding flux from the Earth is obtained by replacing
the prefactor of Eq.~(\ref{eqn:neutrinos}) by $15\, {\rm km}^{-2}\, {
\rm yr}^{-1}$.  Here, $m_\chi$ is the mass of the WIMP, $\rho_\chi$ is
the dark matter density and $\sigma$ is the WIMP-nucleon scattering
cross section.  The function $f(\mchi)$ varies over the range $5 \le
f(\mchi) \le 0.5$ over the mass range $ 10 \le \mchi/{\rm GeV} \le
1000$ for the Sun (with a slightly larger range for the Earth), while
the function $\xi(\mchi)$ is in the range $\sim 0.01-0.3$ over the
same mass range.

The factor $N/\Neq$ is the number of WIMPs in the Sun/Earth, and it is
set by the competing effects of WIMP capture and annihilation. If we
assume the capture rate is $\Cc$ and the annihilation rate is $\Ca$,
then the time $t$ evolution of the number $N$ of WIMPs is given by the
differential equation,
\begin{equation}
\label{eqn:riccati} dN/dt = \Cc - \Ca N^2.
\end{equation} 
If both, $\Cc$ and $\Ca$ are constant with time,
and the initial condition is $N(t=0)=0\equiv N_0$, the solution to
this equation is
\begin{equation}
\label{eq:Nt1} N(t) = \sqrt{ \frac{ \Cc }{\Ca}} \frac{ e^{t/\tau} -
\gamma e^{-t/\tau}}{e^{t/\tau}+\gamma e^{-t/\tau}},
     \end{equation}
where
\begin{equation} \gamma \equiv \frac{ 1 - N_0 \sqrt{\Ca / \Cc}}{1 +
N_0 \sqrt{\Ca / \Cc}} \le 1,
\end{equation}
and $\tau = 1 / \sqrt{\Cc \Ca}$ is the equilibration timecale.  After
a time $t \gtrsim \tau$, the number $N$ approaches $ \Neq \equiv N(t
\gg\tau) = \sqrt{ \Cc / \Ca}$, and the annihilation rate $\Gamma_\nu$
becomes equal to (one half) the capture rate, $\Gamma_\nu = \Ca N^2 /
2 = \Cc / 2$.

The equilibration timescale is determined solely by the capture and
annihilation rates $\Cc$ \& $\Ca$, and thus set by the cross section
of the WIMP-nucleon scattering, and the cross section for WIMP
annihilation. Its value is given by (for the Sun)
\begin{eqnarray}
\label{eqn:tau} \tau_\odot &=& 1.9\times10^5 \, {\rm yrs} \, \left[
\left(\frac{\rho_{\chi}}{0.3 {\rm GeV cm}^{-3}} \right) \left( \frac{
\langle \sigma_A v \rangle}{10^{-26} \, {\rm cm}^{-3}{\rm s}^{-1}}
\right) \right]^{-1/2} \nonumber \nonumber \\ &\times&
\left(\frac{m_\chi}{100\,{\rm GeV}}\right)^{-3/4}
\left(\frac{\sigma}{10^{-40} {\rm cm}^{2}} \right)^{-1/2},
\end{eqnarray} 
while the corresponding value for the Earth can be
obtained by replacing the prefactor of Eq.~\ref{eqn:tau} by $1.1
\times 10^8 {\rm yr}$.  In Eq.~\ref{eqn:tau} $\langle \sigma_A v
\rangle$ is the thermally-averaged annihilation cross section (times
relative velocity $v$ in the limit $v \rightarrow 0$). Typically, the
equilibration timescale for the Sun is smaller than the age of the
Solar System. The equilibration timescale for the Earth is generally
much higher mainly due to differences in mass and composition. It
should be noted that in general, equilibration timescales may vary by
numerous orders of magnitude, not only due to the current
uncertainties in the WIMP physics, but also due to some potentially
more exotic physics such as Sommerfeld \cite{Hisano:2004ds} or
self-capture \cite{Zentner:2009is} enhancements.

The rate in a direct detection experiment is given by,
\begin{eqnarray}
\label{eqn:direct} R_{\rm DD}^{\rm sc} &=& 2.2\times10^5 \, {\rm
kg}^{-1}\,{\rm yr}^{-1} \, \left( \frac{\rho_\chi}{ 0.3 {\rm GeV
cm}^{-3}} \right) \eta_c(m_\chi,m_i) \left(\frac{\mchi}{100\,{\rm
GeV}}\right) \nonumber \\ &\times& \left( \frac{m_i}{100\,{\rm GeV}}
\right) \left( \frac{m_i}{\mchi + m_i} \right)^2
\left(\frac{\sigma}{10^{-40} {\rm cm}^{2}} \right),
\end{eqnarray}
where $m_i$ is the mass of the target nucleus, and $\eta_c(m_\chi,m_i)
$ (given in Ref.~\cite{Griest:1988ma}) accounts for form-factor
suppression.  Comparison between Eqs.~\ref{eqn:neutrinos} \&
~\ref{eqn:direct} shows that a measurement of the two rates gives a
measure of the quantity $N/\Neq$. If the dark matter halo density is
constant with time, and the equilibration timescale is shorter than
the age of the Solar System, then $N = \Neq$ and the two rates are
proportional to each other (modulo the microphysics that describes
each process). If on the other hand the dark matter halo density 
is time dependent (for example the dark matter density was higher 
at some time in the past), then the neutrino signal will be elevated 
relative to the direct detection signal.

\subsection{Time-dependence of the capture rate} If there is Galactic
substructure at the solar Galactic radius ($\sim 8.5 {\rm kpc}$), then
the dark matter density at the position of the Solar System varies
with time. Due to the time lag between capture and annihilation, the
current energetic neutrino flux is determined not by the present local
dark matter density, but rather the density of dark matter along the
past trajectory of the Solar System. On the other hand, the current
direct detection rates in a direct detection experiment are determined
by the present local dark matter density. In general, the behaviour of 
the energetic neutrino signal can be obtained by numerically solving 
Eq.~\ref{eqn:neutrinos} with a time dependent capture rate. Instead, we 
can glean the behaviour of time dependence of the capture rate by 
considering some simple instructive toy examples. 
\begin{figure}
\label{fig:fig2}
  \includegraphics[height=.35\textheight]{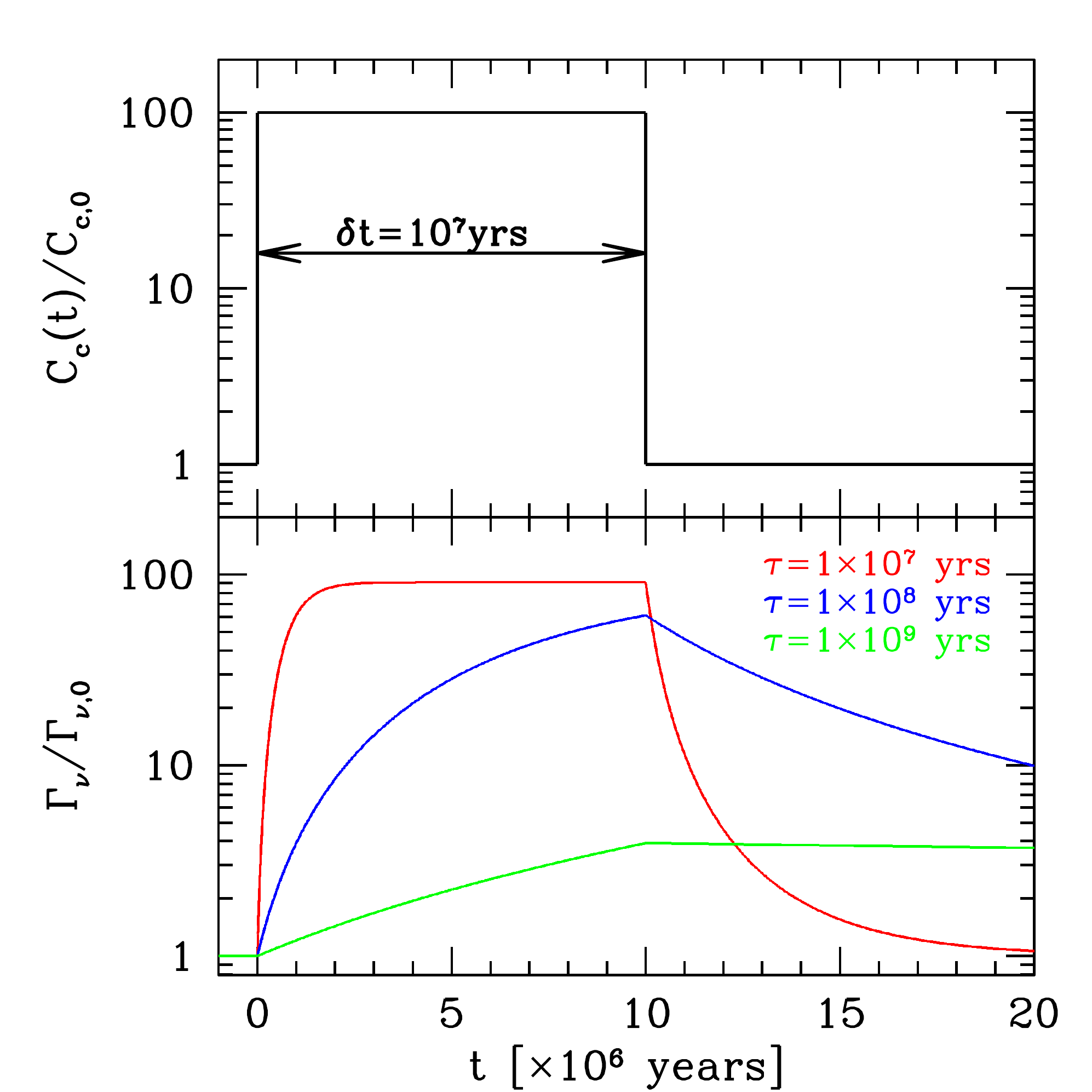}
  \includegraphics[height=.35\textheight]{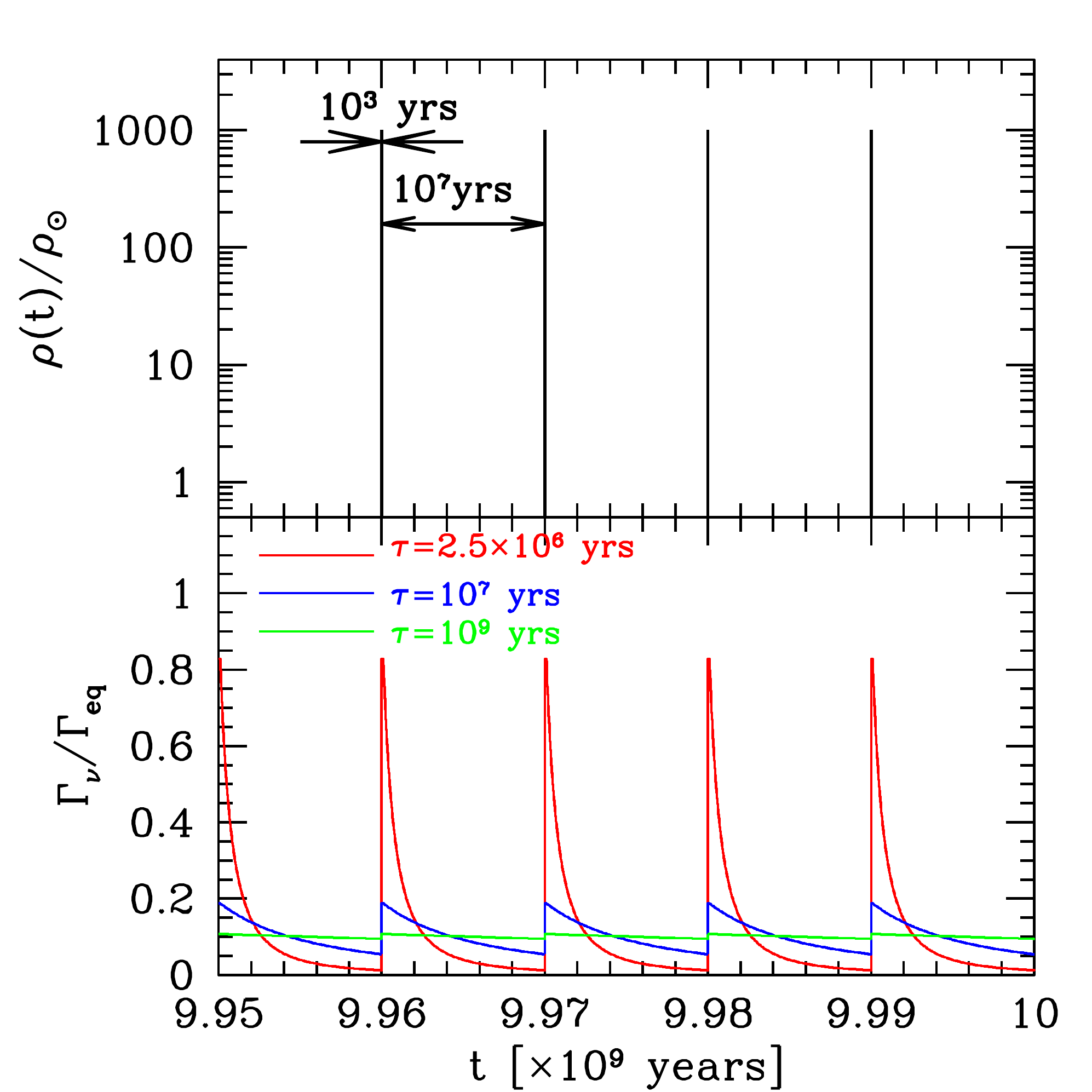}
  \caption{{\it Left}: The effect of a time dependent capture rate to
the energetic neutrino signal from the Sun (red curve) and the Earth
(green curve). The top panel shows the capture rate (proportional to
the direct detection rate). The bottom panel shows the energetic
neutrino flux for three different values of the equilibration
timescale. {\it Right}: The effect of multiple interactions. Line
types same as figure on left. Both figures adapted from
\cite{Koushiappas:2009ee}.}
\end{figure}

Consider the passage of
the Solar System through a region where the density is enhanced by a
factor of $\sim 100$ relative to the smooth density at the Solar
Galactic radius, and that the duration of this enhancement is of order
$\sim 10^7$ years (e.g., a $10^9 M_\odot$ halo of constant
density). The left panel of Fig.~\ref{fig:fig2} shows the effects on
the capture rate, direct detection rate and energetic neutrino signal
for this toy interaction. The direct detection rate is directly
proportional to the dark matter density and thus traces the capture
rate. However, the energetic neutrino signal enhancement is not
instantaneous, but rather is set by the equilibration timescale. Short
equilibration timescales relative to the duration of the capture rate
enhancement in this example (e.g. the Sun) result in a fast neutrino
enhancement, while long equilibration timescales (e.g. the Earth)
result in a lagging signal enhancement. After exiting such a region of
capture rate enhancement, the direct detection signal (capture rate)
will return to the canonical value, the signal from the Sun will
return to the canonical value at some later time set by the short
equilibration timescale, but the signal from the Earth will remain
elevated. A measurement of such elevated signal will then point to a
recent passage through a high-density region of dark matter in the
recent past of the Solar System's history.

It is interesting to consider the case where there are multiple
interactions with regions of enhanced capture rate (e.g., passage
through numerous small scale subhalos). If the timescale between
successive encounters is shorter than the equilibration timescale of
the Sun/Earth, then it is possible the signal of energetic neutrinos
to remain elevated at all times. For the purpose of illustration, the
right panel of Fig.~\ref{fig:fig2} shows the effect of having multiple
successive signal enhancements that would correspond to the presence
of $1 M_\odot$ objects at the solar Galactic radius \cite{earthmass}
(for example if all the local dark matter density was in objects of
mass $1 M_\odot$). The mean timescale between interactions would be
$\sim 10^7 $ years, the duration of each interaction would be $\sim
10^3$ years, and the corresponding capture rate enhancement would be
$\sim 1000$. In this case, for equilibration timescales which are of
order the mean timescale between encounters, the energetic neutrino
signal is depleted completely prior to the next encounter, while for
longer equilibration timescales, the net effect is an elevated signal
at all times. In terms of the signals from the Sun and the Earth, this
means that the signal from the Earth will be boosted relative to the
signal from the Sun for most of the time.
\begin{figure}
\label{fig:fig3}
  \includegraphics[height=.35\textheight]{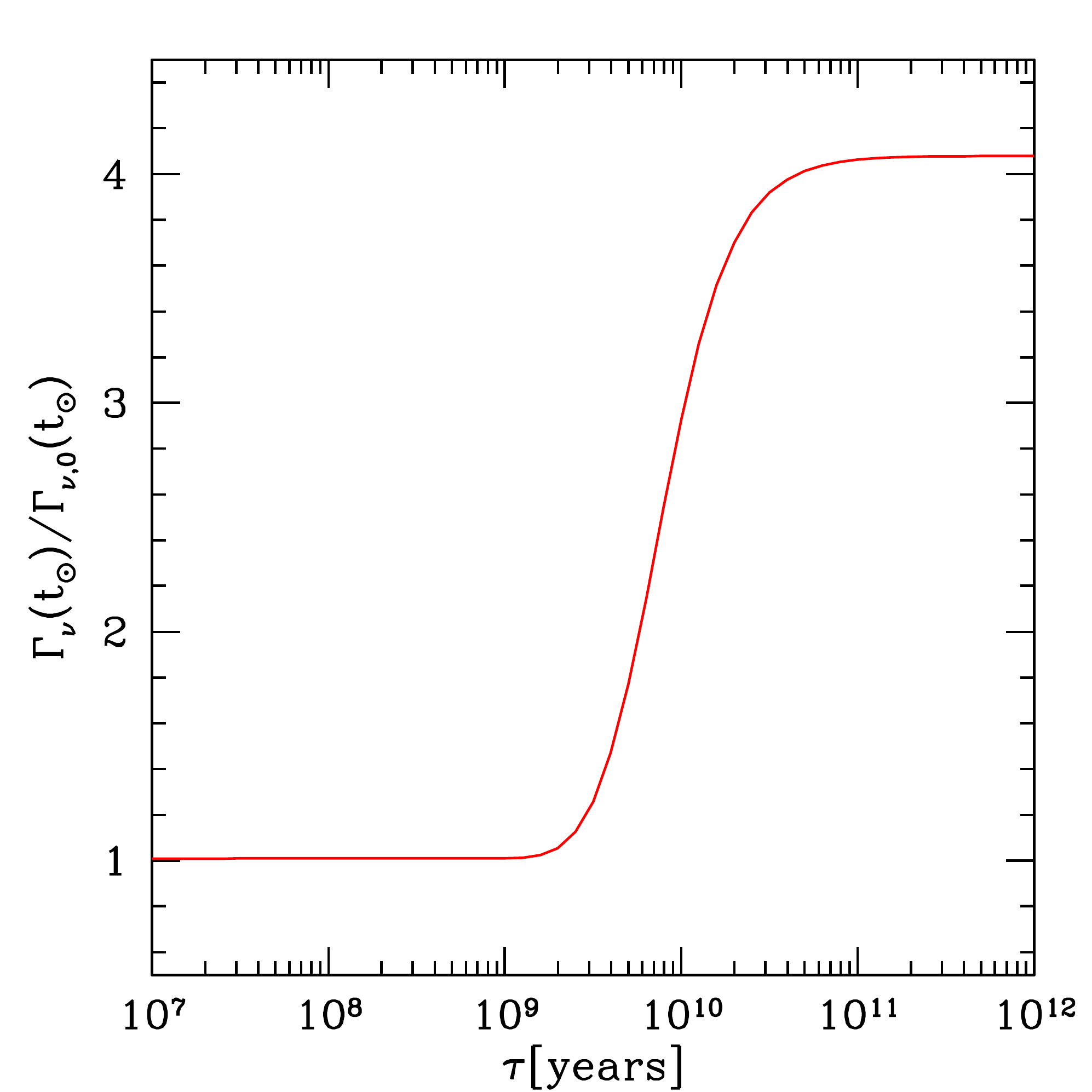}
  \caption{The net effect of the presence of a population of $10^{-6}
M_\odot$ objects in the Milky Way halo. For long equilibration
timescales, the energetic neutrino signal is higher than the signal
that would be obtained from a smooth-only dark matter distribution.}
\end{figure}

The presence of substructure effectively speeds up equilibration 
if the equilibration timescale is larger than the age of the solar 
system. For example, suppose that the dark matter has some smooth
component, and a substructure population that extends all the way down
to the smallest scales that correspond to cold dark matter particles \cite{earthmass},
consistent with an extrapolation of the substructure mass function
found in numerical simulations \cite{Springel:2008cc,Diemand:2008in}.
If we only consider objects of mass $10^{-6} M_\odot$, i.e., a mean density
enhancement of $\sim 100$ with duration of 50 years, and a frequency
of one per million years, then for long equilibration timescales
(longer than the age of the Solar System) the amount of depletion of WIMPs is
negligible relative to the amount of WIMPs being captured. 
This effect leads to a continuous buildup of WIMPs, i.e.,
an energetic neutrino signal that today is higher than the signal that
would be obtained if the Milky Way halo was
smooth. Fig.~\ref{fig:fig3} shows this effect as a function of the
equilibration timescale. Long equilibration timescales result in a net
build-up of WIMPs. Therefore, as $N$ approaches $\Neq$ faster over a
given timescale (e.g., the age of the solar system) due to
substructure, the equilibration of a WIMP population is sped up.

\section{Conclusions} 
We show that the presence of substructure in the
Milky Way halo alters the energetic neutrino signal expected from the
Sun and the Earth. A future measurement of the direct detection rate
in dark matter experiments, and the rate of energetic neutrinos from
the Sun/Earth can be used as a probe of the amount of substructure in
the Milky Way. In addition, the comparison between the energetic
neutrino signals from the Sun and the Earth can yield information on
the recent past history of dark matter density at the solar Galactic
radius and the Solar System's dark matter encounters in its past.

%---------------- ACKNOWLEDGEMENTS

\begin{theacknowledgments} SMK thanks The Ohio State University's Center for
Cosmology and AstroParticle Physics (CCAPP) for hospitality during the
CCAPP Symposium. \end{theacknowledgments}

%---------------- BIBLIOGRAPHY

%---------------- END
\end{document}